\documentclass[aps,prd,preprint,superscriptaddress,showpacs]{revtex4}
\usepackage{epsf,epsfig,graphics,graphicx}
\usepackage{verbatim,color,ulem}
\newcommand{\be}{\begin{equation}}
\newcommand{\ee}{\end{equation}}
\newcommand{\bea}{\begin{eqnarray}}
\newcommand{\eea}{\end{eqnarray}}
\newcommand{\ba}{\begin{array}}
\newcommand{\ea}{\end{array}}
\usepackage{amssymb,amsmath,amsthm,graphicx}
\usepackage[mathscr]{eucal}
\usepackage{enumerate,color,verbatim,multirow,comment}

\begin{document}
\title{Environment-induced uncertainties on moving mirrors in quantum critical theories via holography}

\author{Da-Shin Lee}
\email{dslee@mail.ndhu.edu.tw} \affiliation{Department of Physics,
National Dong-Hwa University, Hualien, Taiwan, R.O.C.}

\author{Chen-Pin Yeh}
\email{chenpinyeh@mail.ndhu.edu.tw} \affiliation{Department of
Physics, National Dong-Hwa University, Hualien, Taiwan, R.O.C.}

\begin{abstract}
Environment effects on a $n$-dimensional mirror from the strongly
coupled d-dimensional quantum critical fields with a dynamic
exponent $z$ in weakly squeezed states are studied by the
holographic approach. The dual description is a $n+1$-dimensional
probe brane moving in the $d+1$-dimensional asymptotic Lifshitz
geometry with gravitational wave perturbations. Using the
holographic influence functional method, we find that the large
coupling constant of the fields reduces the position uncertainty
of the mirror, but enhances the momentum uncertainty. As such, the
product of the position and momentum uncertainties is independent
of the coupling constant. The proper choices of the phase of the
squeezing parameter might reduce the uncertainties, nevertheless
large values of its amplitude always lead to the larger
uncertainties due to the fact that more quanta are excited as
compared with the corresponding normal vacuum and thermal states.
In the squeezed vacuum state, the position and momentum of the
mirror gain maximum uncertainties from the field at the dynamic
exponent $z=n+2$ when the same squeezed mode is considered. As for
the squeezed thermal state, the contributions of thermal
fluctuations to the uncertainties decrease as the temperature
increases in the case $1<z<n+2$, whereas for $z>n+2$ the
contributions increase as the temperature increases. These results
are in sharp contrast with those in the environments of the
relativistic free field. Some possible observable effects are
discussed.

\end{abstract}


\maketitle
\section{Introduction}
Macroscopic quantum phenomena often refer to collective quantum
behavior in objects, consisting of a large number of particles in
atomic scales~\cite{Leggett_02,Caldeira_book}. The best known
examples are superconductivity and superfluidity. Additionally,
experimental realizations of Bose-Einstein condensation in dilute
gases certainly provide a more fruitful venue, in which various
macroscopic quantum phenomena are explored under experimental
controls. Moreover the progress in electro- and opto-mechanical
techniques makes it possible to prepare macroscopic or mesoscopic
mechanical objects in nearly pure quantum
states~(See~\cite{Schwab,Aspelmeyer,Armour,Arndt}) where the
center of mass of a object obeys a quantum mechanical equation of
motion. Recently experiments to demonstrate quantum interference
between the macroscopic objects have been proposed
in~\cite{Marshall,Muller}. In those experiments it is essential
that a macroscopic system like the mirror is prepared in the
quantum superposition state.

Because of a large number of the degrees of freedom in
macromechanical systems, the observability of the quantum behavior
will be strongly influenced by interactions with the environment
and the experimentally accessible quantum region will also depend
on the decoherence dynamics due to the presence of the
environment~\cite{Armour}. A viable microscopic approach to
investigate the environmental effects on the system would start
with a specific system-environment model. Then the environmental
degrees of freedom are integrated out by the method of
Feynman-Vernon influence functional. This approach consistently
and systematically accounts for the influence of the environment
on the system of interest~\cite{Leggett,SK,Fv}. The influence functional can
be exactly derived if the environment variables are Gaussian and
their coupling with the system is linear~\cite{GSI,Hu}. In
particular, the  effects from the quantized electromagnetic fields
on a point charge in the dipole approximation have been studied
extensively by~\cite{Lee_06,Lee_08,Lee_09,Lee_12}.

In the work~\cite{Unruh_89}, the environment is modeled by a free
massless scalar field in vacuum and thermal states, and its
coupling to the system of the particle, which is a harmonic
oscillator, is linear in particle's position. They focused on the
evolutions of particle's reduced density matrix which initially is
in vacuum and squeezed states, and explored the uncertainties of
particle's position and momentum due to the interaction with the
environment. What they found is that if the system is prepared in
a pure state, the loss of quantum coherence can happen as a result
of the coupling to the environment. In particular, when the
environment field is in zero temperature, the off-diagonal terms
of the reduced density matrix in the position representation
decrease more rapidly than in the momentum representation,
resulting in relatively small position uncertainty. This comes
from the fact that the system is coupled to the environment by its
position variable. They also discussed the changes in these
uncertainties by varying the squeeze parameters of the system and
the temperature of the environments. Here we would like to explore
these effects from the environments of strongly coupled fields and
also allow the dimensions of probe objects and environments to be
arbitrary. The purpose is to make possible comparisons with
various cases in weakly coupled environments.

In quantum field theory, the correlators of weakly interacting
quantum fields are normally computed perturbatively in terms of
the small coupling constant. As for strongly coupled fields in
high dimensions, the holographic correspondence is among very few
known nonperturbative ways to calculate their correlators. Thus in
this paper, we will extend the results in~\cite{Unruh_89} by
considering the strongly coupled environment that admits a
holographic description.  The idea of holographic duality is
originally proposed as the correspondence between $4$-dimensional
conformal field theory (CFT) and  gravity theory in
$5$-dimensional anti-de Sitter (AdS) space~\cite{AdSCFT}. Other
backgrounds and field theories are soon to be generalized with the
possibility to study the strong coupling problems in the condensed
matter systems (see \cite{Hartnoll_09} for a review). Considerable
efforts also have been focused on using the holography idea to
explore the Brownian motion of a particle moving in a strongly
coupled
environment~\cite{Herzog:2006gh,Gubser_06,Teaney_06,Son:2009vu,Giecold:2009cg,CasalderreySolana:2009rm,Huot_2011,Holographic
QBM,Tong_12,Hartnoll_10,Gusber_08,Gursoy_10,Kovtun_05,Kiritsis_13,Rajagopal_15,Roy_15,Ban_14,mirror,Yeh_14,Yeh_16_1,Yeh_16_2}. A review
on the holographic Brownian motion can be found in
\cite{Holographic QBM}.

Here we will apply a bottom-up holographic method, proposed in our
earlier work~\cite{mirror}, to find the uncertainties of a
$n$-dimensional mirror in the  environment of $d$-dimensional
quantum critical theories at zero and finite temperature. The
holographic dual for such quantum critical theories has been
proposed in~\cite{Kachru_08} where the gravity theory is in the
Lifshitz background (See~\cite{Tong_12,Hartnoll_10} for details).
Several physical phenomena have been studied in this theory,
including linear DC conductivity, power-law AC conductivity, and
strange fermion
behaviors~\cite{Hartnoll_10,Gursoy_12,Alisha,Gursoy,dimer model}.
In our set-up, the bulk counterpart of the mirror is a
$(n+1)$-brane in the Lifshitz geometry in $d+1$ dimensions. The
motion of the mirror can be realized from the dynamics of the
brane at the boundary of the bulk. As explained in \cite{mirror}
and will also be reviewed in the appendix, this holographic
identification is based upon the fact that the coupling of the
brane to the boundary field shares similar feature as the coupling
between the mirror and the environment quantum field where the
mirror is of perfect reflection for the field~\cite{wu,mirror}.
The force on the mirror is given by the position change of the
mirror~\cite{mirror}. It has been discussed in~\cite{mirror} that,
for $z = 1$ (the relativistic environmental field) and for a
2-dimensional mirror, the ohmic dynamics of the mirror with the
$T^4$ dependence of the damping constant due to the strongly
coupled environment field at  finite temperature $T$, is in
agreement with the finding in~\cite{wu} where the  relativistic
thermal free  field  is considered.  It is also
expected that the proportionality constant in the damping constant
is different between the strongly coupling environment field and the
free field. In this paper, we consider the environment
of the squeezed vacuum (thermal) baths, whose holographic duals
arise from gravitational wave perturbations in Lifshitz (black
hole) background, as suggested
in~\cite{Lenny_99,Yeh_16_1,Yeh_16_2}. Using the method of
holographic influence functional, developed in~\cite{Yeh_14}, we
then study the uncertainties of the position and momentum of the
mirror.

Our presentation is organized as follows. In next section, we
briefly review the method of holographic influence functional for
the squeezed states and explain the duality between the squeezed
state and the gravitational wave perturbed Lifshitz black hole.
The reviews on the method of influence functional in field theory
and the construction of holographic influence functional for pure
Lifshitz geometry and Lifshitz black hole are in the appendices.
In Sec.~\ref{sec3}, the evolution of the uncertainties of a
mirror, influenced by the environment of strongly coupled fields,
are computed.  The comparisons with the results from the
environment given by the relativistic free-field are also
discussed. Summary and outlook are in Sec.~\ref{sec4}.

The sign convention $(-,+,....,+)$ is adopted in the $d +
1$-dimension metric in dual gravity theory with indices
$\mu,\nu,..$. Indices $a, b,c,..$ denote all spacetime coordinates
in the boundary field theory while $i,j, k...$ denote only spatial
dimensions.

\section{{Holographic influence functional for the environment field in squeezed vacuum/thermal states}}
\label{sec1}

The environment fields we consider in this paper is the theory of
quantum critical points with the following scaling symmetry:
  \be\label{E:ernfd}
  t\rightarrow\mu^zt \, ,\qquad\qquad x\rightarrow\mu x \,
  \ee
where $z$ is called the dynamical exponent. The holographic dual
for such quantum critical theories in 2+1-dimension has been
proposed in~\cite{Kachru_08}, where the corresponding gravity
theory is in the 3+1-dimensional Lifshitz background. Here we consider the d+1-dimensional background, which is
asymptotic to the Lifshitz metric,
   \be
   \label{bmetric}
   ds^2=g^{(0)}_{\mu\nu}dx^{\mu}dx^{\nu}=-\frac{r^{2z}}{L^{2z}}dt^2+\frac{1}{r^2}dr^2+\frac{r^2}{L^2}dx_idx_{i}
   \, ,
   \ee
where the scaling symmetry {\eqref{E:ernfd}} is realized as an
isometry of this metric. This d+1-dimensional Lifshitz metric can
be constructed by coupling gravity with negative cosmological
constant to massive Abelian vector fields \cite{Marika_08}. The
corresponding action is given by:
  \be
  \label{action}
  S=\frac1{16\pi G_{d+1}}\int
  d^{d+1}x \,\sqrt{-g}\, ( R+ 2\Lambda-\frac14 {\cal F}^{\mu\nu}{\cal F}_{\mu\nu}-\frac1 2m^2 {\cal A}^{\mu} {\cal
  A}_{\mu})\,.
  \ee
The action yields the equations of motion for the metric and the
vector fields,
   \bea \label{EoM}
   &&R_{\mu\nu}=-\frac{2 \Lambda}{d-1} g_{\mu\nu}+\frac12 g^{\alpha\beta}{\cal F}_{\mu\alpha} {\cal F}_{\nu\beta}+\frac12 m^2 {\cal A}_{\mu} {\cal A}_{\nu}-\frac1{4(d-1)} {\cal F}_{\alpha\beta} {\cal F}^{\beta\alpha}g_{\mu\nu}\, ,\nonumber\\
   &&D_{\mu}{\cal F}^{\mu\nu}=m^2 {\cal A}^{\nu} \, ,
   \eea
where $D_{\mu}$ is the covariant derivative. The solutions of the
vector fields are assumed to be
  \be \label{A_field}
  {\cal A}_{\mu}={\cal A} \frac{r^z}{L^z}\delta^{0}_{\mu} \, .
  \ee
Then the Lifshitz background in (\ref{bmetric}) is {the solution
of (\ref{EoM}) with}
 \be
 \label{para}
 {\cal A}=\sqrt{\frac{2(z-1)}{z}},~~m^2=\frac{(d-1)z}{L^2},~~\Lambda=\frac{(d-1)^2+(d-2)z-z^2}{2L^2}
 \, .
 \ee
In particular, we consider the Lifshitz black brane perturbed by
the gravitational wave with metric (we set the radius of curvature
$L$ to one),
  \be
  \label{lifshitz bh gw}
  ds^2=-r^{2z}f(r)dt^2+\frac{dr^2}{f(r)r^2}+r^2 dx_i dx_i+r^2\phi(t,r)\xi_{\mu\nu}dx^{\mu}dx^{\nu} \, ,
   \ee
where $\xi_{\mu\nu}$, the polarization tensor, has non-zero
components only in the spatial directions $(i,j)$ of the boundary,
and is assumed transverse and traceless. Here $\phi (t,r)$ is
assumed to be small, and its equation of motion will be determined
later. We also have $f(r)\rightarrow1$ for $r\rightarrow\infty$
and $f(r)\simeq c(r-r_h)$ near the black brane horizon $r_h$ with
$c=({d+z-1})/{r_h}$. For example, the model in (\ref{action}), for
the case $d=3$ and $z=2$, has the exact black hole solution
\cite{Peet_09},
 \be
 f(r)=\sqrt{1+\frac1{10r^2}-\frac{3}{400r^4}}
 \ee
Also, in the case of $d=3$ and $z=1$,  the AdS black brane
solution is found with $f(r)=1-\frac{r_h^3}{r^3}$. However, for
general values of $z$ and $d$, only numerical and perturbative
solutions are available. Nevertheless, the detailed form of $f(r)$
is not very relevant when considering the perturbations in low
frequency limits, as will be seen in our subsequent discussions.
The black hole temperature, which is also the temperature in the
boundary field theory, is
 \be
 \label{BHT}
\frac1T=\frac{4\pi}{d+z-1}\frac1{r_h^z} \, .
  \ee
As suggested in \cite{Lenny_99} and also in our earlier
works~\cite{Yeh_16_1, Yeh_16_2} a possible holographic realization
of the weakly squeezed vacuum (thermal) state of the boundary
field is given by gravitational wave perturbed Lifshitz (black
brane) background in (\ref{lifshitz bh gw}). To justify this
identification, we consider two ways of deriving correlators for
the position of mirror. One is by the holographic influence
functional method \cite{Yeh_14} and the other is through the
Bogoliubov transformations of the excitations on the probe brane.
We first describe the holographic influence functional in the
following.

In quantum field theory, the influence functional is a way to
summarize the effect of the quantum field to a mirror's position.
We give a review in Appendix~\ref{inf}. In the holographic setup,
the dual description of the mirror is a $n+1$-dimensional probe
brane in the asymptotic Lifshitz background (\ref{lifshitz bh
gw}). In accordance with the closed-time-path
formalism~\cite{Leggett,SK,GSI} that we have discussed in
Appendix~\ref{inf}, we introduce $Q^+(t,r_1)$ and $Q^-(t,r_2)$,
which correspond to the branes living in two regions with
different asymptotic boundaries in the maximally extended Lifshitz
black hole geometry~\cite{Son_09, Son_02}. $Q^+(t,r_1)$ and
$Q^-(t,r_2)$ are determined by their analytical properties at
$r=r_h$ \cite{Son_02} or equivalently by the unitarity arguments
of the boundary theory~\cite{Yeh_14}. We also impose the following
boundary conditions
 \be
 \label{bc} {X^{\pm}(t)=Q^{\pm}(t,r_{b})}\,,
 \ee
{where the variable $X (t)$} can be identified as the displacement
of the moving mirror. Then the classical on-shell action of the
brane is identified as the influence functional for the
mirror~\cite{Son_09}:
   \be
   \label{gravity action}
   \mathcal{F} [q^+,
   q^-]= S_{gravity}\left(Q^+(t,r_b),Q^-(t,r_b)\right) =S^{\rm on-shell}_{DBI}(Q^+)-S^{\rm on-shell}_{DBI}(Q^-)\,
   \ee
where $S^{\rm on-shell}_{DBI}$ is the on-shell DBI action for the
probe brane. To quadratic order with the background~(\ref{lifshitz bh gw}) , we can write
 \be
  S_{DBI} =- \frac{T_{n+1} S_n}{2}\int dr \, dt \,
\bigg( r^{z+n+3} \,f(r) \,(1+ \phi (r, t)) \,  X^{I'}
 X^{I'} -  \,(1+ \phi (r, t)) \, \frac{{\dot X}^I  {\dot X}^I}{f(r) {r^{z-n-1}}}\bigg)
\, , \label{S_DBI}
  \ee
where $T_{n+1}$ and $S_n$ are the tension and area of the brane
respectively and $X^I(t,r)$ parameterizes the brane's position
with $I=n+1,...,d$ denoting the transverse directions to the brane.
Also, $X'^{I} =\partial_{r} X^{I}$, $\dot{X}^{I}=\partial_{t}X^{I}$.
We assume the mirror does not deform when moving in its transverse
directions so $X^{I}$'s depend only on $t$ and $r$. Then, up to the
first order in $\phi$, the equation of motion of the perturbation
of the brane derived from~(\ref{S_DBI}) becomes
 \bea
  \label{DBI with Ts}
&& \frac{\partial}{\partial r}\biggl[ r^{z+n+3}\, f(r)
\frac{\partial }{\partial
 r} X^I (r, t) \biggr]-\frac{\partial}{\partial t}\, \biggl[ \frac{1}{f(r) \, r^{z-n-1}}  \,
 \frac{\partial}{\partial t} X^I (r, t)\biggr] \nonumber\\
 &&\quad\quad =-\frac{\partial}{\partial r}\biggl[f(r) r^{z+n+3}\,  \, \phi (r, t)\, \frac{\partial }{\partial
 r} X^I (r, t) \biggr]+\frac{\partial}{\partial t}\, \biggl[ \frac{1}{f(r) \, r^{z-n-1}}  \, \phi (r, t)\,
 \frac{\partial}{\partial t} X^I (r, t)\biggr]
 \, .
  \eea
Using the equation of motion above, the classical on-shell action
with the boundary terms is given by \be \label{S_T_DBI_onshell}
S_{DBI\, (1+\phi)}^{{\rm on-shell}} \simeq -\frac{T_{n+1} S_n}{2}
r_b^{z+n+3} \int dt \, (1+\phi(t,r_b)) \left( X^I (t,r_b)
\partial_r X^I (t,r_b)
\right)\, .
  \ee
The solution in frequency space can be expressed perturbatively as
  \be
   X^I_{\omega}(r)=X^{I(0)}_{\omega}(r)+X^{I (\phi)}_{\omega}(r) \, ,
  \ee
where the zeroth-order solutions $X^{I(0)}_{\omega}(r)$ at zero temperature and finite
temperature will be reviewed in Appendix~\ref{zero order} and are
given respectively by \eqref{mode} for all $\omega>0$ and
\eqref{modeT} for small $\omega$. Then the equation of motion for
$X^{I (\phi)}_{\omega}(r)$ to leading order is given by
  \bea
  \label{PertuEoM}
  &&\frac{\partial}{\partial r}
  \bigg[ r^{z+n+3}\, f(r) \partial_rX^{I(\phi)}_{\omega}(r)\bigg]+ ( r^{-z+n+1}/f(r)) \, \omega^2X^{I(\phi)}_{\omega}(r)\nonumber \\
  && \quad=- \int
  d\omega'\,  \bigg[ r^{z+n+3}\,f(r)\, \partial_r\phi (\omega, r)\, \partial_r
  X^{I(0)}_{\omega-\omega'}(r)  +\omega'(\omega-\omega')\, (r^{-z+n+1}/f(r)) \,  \phi (\omega', r)\, X^{I(0)}_{\omega-\omega'}(r)
  \bigg] \, . \nonumber\\
  \eea
Now we would like to find the perturbed on-shell action.  To this
end, we only need the asymptotical forms of the solutions
$\phi(t,r)$ and $X^{I(\phi)}_{\omega}(r)$  in the large $r$ limit
by taking the $f(r) \rightarrow 1$ limit in~(\ref{PertuEoM}). In
addition, the dynamics of $\phi(t,r)$ in the limit of large $r$
can also be obtained by considering the gravitational waves on the
Lifshitz metric $g_{\mu \nu}^{(0)}$ in~(\ref{bmetric}). Thus, the
equation of motion for $\phi(t,r)$ in this limit is
   \be
   \label{eom}
 {r^{-2z}}\, \frac{\partial^2}{\partial t^2} \phi(t,r)+(d-1+z)\, r\frac{\partial}{\partial r}\phi
(t,r)+r^2\,
   \frac{\partial^2}{\partial r^2} \phi(t,r)=0 \, ,
   \ee
which can be derived by linearizing~(\ref{EoM}) about the
background solutions~(\ref{bmetric}) and~({\ref{A_field}}). The
Fourier transform of the $\phi(t,r)$ field in frequency space is
defined as
  \be
  \label{phi}
  \phi (t,r)= \int_{0}^{\infty}
  \frac{d\omega}{ 2\pi}\, \phi (\omega,r)   \, e^{-i \omega t}\, ,
  \ee
and the normalizable solution of (\ref{eom}) is
 \be
  \label{varphi}
 \phi (\omega,r)  =r^{-\frac{d+z-2}{2}} \,  \varphi (\omega) \, J_{\frac{d+z-2}{2z}}\bigg(\frac{ \omega}{zr^z}\bigg) +c.c.  \, .
  \ee
The function $\varphi (\omega)$ is determined by the boundary
conditions of the gravitation waves at $r=r_b$. Note that we
impose the normalizable boundary condition for $\phi(t,r)$ at
$r_b$ rather than the infalling boundary condition in the black
brane horizon, normally adopted to construct the retarded
correlators. This boundary condition is obtained by adding the
correct boundary counterterms in the gravity theory so that the
dual boundary stress tensor obtained satisfies the trace Ward
identity and is independent of $r_b$\cite{Kraus_99,Ross_09}. This
is also necessary for us to identify $\varphi(\omega)$ as a
squeezed parameter in the boundary theory \cite{Yeh_16_2}.

The asymptotical form of the solution $X^{I(\phi)}_{\omega}(r) $
in the large $r$ limit has been discussed in
\cite{Yeh_16_1,Yeh_16_2}, where we found that, to the leading
order in $\varphi(\omega)$, we can ignore the contribution of
$X^{I(\phi)}_{\omega}(r) $ in the zero temperature on-shell action
by taking the large $r_b$ limit. Here the same conclusion can be
reached for the finite-temperature on-shell action
(\ref{S_T_DBI_onshell}) due to the similar arguments as follows.
Using (\ref{PertuEoM}), the dependence of $X^{I(\phi)}_{\omega}(r)
$ on $r$ for large $r$ is mainly determined by $\partial_r [
r^{z+n+3}
\partial_r X_{\omega}^{I(\phi)} (r) ] \approx r^{-z+n+1}
\phi  (\omega, r)$ as $f(r_b) \rightarrow 1$ and
$X^{I(0)}_{\omega}(r_b) \rightarrow 1 $ in~(\ref{mode}). From the
asymptotic behavior of $\phi (\omega,r) \approx r^{-d-z+2}$ as
given in~(\ref{varphi}), it leads to $ X_{\omega}^{I(\phi)} ( r)
\approx {r^{-d-3z+2}}$. Thus we conclude that $\phi (\omega, r_b)
>> X_{\omega}^{I(\phi)} (r_b)/ X_{\omega}^{I(0)} (r_b)$ for large
$r_b$, so the contributions from $X_{\omega}^{I(\phi)}$ to the
above perturbed action \eqref{S_DBI} can be ignored when we keep
terms up to linear order in $\varphi(\omega)$. Thus, using the
holographic influence functional prescription~(\ref{gravity
action}) where $Q$  denotes one of the directions $X^I$,  the
on-shell perturbed action $S^{{\rm on-shell}}_{DBI \, \phi}$
obtained from~(\ref{S_T_DBI_onshell}) is expressed as
 \bea \label{S_phi} && S^{\rm on-shell}_{DBI \, \phi}(Q^+)-  S^{\rm on-shell}_{DBI \, \phi}(Q^-) = \nonumber\\
  &-& \frac{T_{n+1} S_n}{2} r_b^{z+n+3}\int dt \, \phi (t,r_b) \, \left( Q^+ (t,r_b)
\partial_r Q^+ (t,r_b)-  Q^-
(t,r_b) \partial_r Q^- (t, r_b) \right) \, \nonumber\\
&=& -\frac{T_{n+1} S_n}{2} r_b^{z+n+3} \int \frac{d\omega}{2\pi}
\int \frac{d \omega'}{2\pi}\, \phi(\omega+\omega',r_b)\left(
Q^+_{-\omega} (r_b) \,
\partial_r Q^+_{-\omega'} (r_b)- Q^-_{-\omega} (r_b) \,
\partial_r Q^-_{-\omega'} ( r_b)\right)\, .  \nonumber\\
 \eea
Substituting the expression (\ref{Q_pm}) with $
\mathcal{X}_{\omega}(r)$ constructed by the zeroth-order solutions
$X^{I(0)} (r)$ in Appendix~\ref{zero order} for zero and finite
temperature into the above expression, the respective perturbed
holographic influence functionals are obtained. These
nonequilibrium Green's functions constructed from the perturbed
influence functionals can be compared with the form of the Green's
functions in the squeezed vacuum and thermal states, obtained by
means of the corresponding squeeze operator as we will describe in
the following. Then the function $\varphi$ in~(\ref{varphi}),
determined by the boundary condition of gravitation waves, can be
identified as the squeezing parameters of the squeezed vacuum and
thermal states.

Now we turn to the second way of deriving correlators for the
mirror's position. According to the holographic
correspondence~\cite{Holographic QBM,Tong_12}, the correlation
functions of the boundary fields can also be found from the
correlation functions of the probe brane's position by taking the
near boundary limit. From the boundary point of view, this link
between two ways of deriving correlation functions is established
through the Langevin equation for mirror's position, which can
also be derived from the corresponding holographic influence
functional~\cite{Yeh_16_1}. Using this method, we first consider
the mode expansion of the brane's position operator ${X} (t,r).$
As in previous discussions, we assume that the brane's position is
independent of $x_1, x_2,...,x_n$. In the small displacement
limit, the dynamics of the brane in different directions along
$X^I$ with $I = n + 1,...,d$, are decoupled. Thus we can just
consider the brane's motion in one of those directions, which is
denoted by ${X} (t,r)$ . Accordingly, the mode expansion of the
position operator evaluated at $r=r_b$, which is identified as the
position of the mirror is given as
  \be
\label{modeexpand} {\qquad} X(t) \equiv {X} (t,r_b) =
\int_0^{\infty} d\omega \, U_{ \omega} (r_b) \bigg( a_{\omega} \,
e^{-i \omega t} + a^{\dagger}_{\omega} \, e^{i \omega t} \bigg) \,
,
  \ee
where $a_{\omega}$ and $a_{\omega}^{\dagger}$ are the annihilation
and creation operators, and they obey  canonical commutation
relations
  \be [ a(\omega), a^\dagger (\omega')
]=\delta(\omega,\omega') \,  , \,\,\, [ a(\omega), a (\omega') ]=[
a^\dagger (\omega), a^\dagger (\omega') ]=0 \, .
  \ee
In the background of Lifshitz black hole, the brane's
perturbations are in thermal states where $\langle
a^{\dagger}_\omega a_{\omega}\rangle_{T}=
(e^{\frac{\omega}{T}}-1)^{-1}$ with black hole temperature $T$.
The mode functions $U_{\omega}(r)$ are found from (\ref{mirror n
with T}) with the Neumann boundary condition and the Wronskian
condition (See~\cite{mirror} for details). Since the mode function near the horizon $r=r_h$
exhibits logarithmic divergence, an infrared energy cutoff scale
as $ r \rightarrow r_h$ is introduced for regularization. We may
absorb this infrared divergence by carefully defining the density
of states~\cite{Holographic QBM}. The square of the
divergence-free mode function in the low frequency limit is
obtained in our previous work~\cite{mirror} as
\begin{equation} \label{UT}
  U^{T \, 2}_{\omega} (r_b)= \frac{1}{\pi \omega T_{n+1} S_n } r_h^{-n-2} +\mathcal{O}(\omega^0)\, .
\end{equation}
In the zero-$T$ limit, the mode function squared evaluated at $r_b$
can be found exactly as~\cite{mirror}
\begin{equation} \label{U0}
   U^{(0)\, 2}_{\omega} (r_b)=\frac{2 z r_b^{z-2-n}}{\pi^2\, \omega^2 T_{n+1} S_n}\frac{1}{J^2_{\frac{n+2}{2z}-\frac{1}{2}} (\frac{\omega}{z r^z})+ Y^2_{\frac{n+2}{2 z}-\frac{1}{2}}(\frac{\omega}{z r^z})}\, .
\end{equation}
The squeezed vacuum and thermal states can be constructed from the
Bogoliubov transformations of the creation and annihilation
operators of the normal vacuum and thermal states. Here we assume
the general two-mode squeezed thermal states. We will see later
that the corresponding Green's functions of boundary fields, in
the small squeeze parameter limit, have the same form as those of
the Green's function constructed from the perturbed influence
functional in gravitational wave perturbed Lifshitz black hole.
The squeezed vacuum state can be obtained by taking the zero
temperature limit. Using the squeeze operator, the two-mode
squeezed thermal states can be defined as
  \be
   \rho^T_{\xi_{\omega\omega'}} =S(\xi_{\omega\omega'})\,\rho_T \, S^\dagger (\xi_{\omega\omega'}) \,,\quad
S(\xi_{\omega\omega'})=\exp\left[\frac{1}{2}\bigl(\xi_{\omega\omega'}^{*}\,a_{\omega}a_{\omega'}-\xi_{\omega\omega'}\,a_{\omega}^{\dagger}a_{\omega'}^{\dagger}\bigr)\right]\,
 \ee
with the thermal density matrix $\rho_T$ in~(\ref{initialcondphi})
and the squeeze parameter
$\xi_{\omega\omega'}=r_{\omega\omega'}\,e^{i\theta_{\omega\omega'}}$.
With the help of the Baker-Campbell-Hausdorff formula, we readily
find the Bogoliubov transformations of the creation and
annihilation operators due to the squeeze operator
$\mathcal{S}\left(\xi_{\omega\omega'}\right)$,\begin{eqnarray}
    &&\mathcal{S}^{\dagger}\left(\xi_{\omega\omega'}\right)a_{\omega}\,\mathcal{S}\left(\xi_{\omega\omega'}\right)=\mu_{\omega\omega'}a_{\omega}-\nu_{\omega\omega'}a^{\dagger}_{\omega'}\,,\qquad\text{and}\qquad\mathcal{S}^{\dagger}\left(\xi_{\omega\omega'}\right)a^{\dagger}_{\omega}\mathcal{S}\left(\xi_{\omega\omega'}\right)=\mu_{\omega\omega'}a^{\dagger}_{\omega}-\nu^*_{\omega\omega'}a_{\omega'}\,,\nonumber\\
&&\mathcal{S}^{\dagger}\left(\xi_{\omega\omega'}\right)a_{\omega'}\,\mathcal{S}\left(\xi_{\omega\omega'}\right)=\mu_{\omega\omega'}a_{\omega'}-\nu_{\omega\omega'}a^{\dagger}_{\omega}\,,\qquad\text{and}\qquad\mathcal{S}^{\dagger}\left(\xi_{\omega\omega'}\right)a^{\dagger}_{\omega'}\mathcal{S}\left(\xi_{\omega\omega'}\right)=\mu_{\omega\omega'}a^{\dagger}_{\omega'}-\nu^*_{\omega\omega'}a_{\omega}\,,
\nonumber \\
\end{eqnarray}
and we have
 \begin{align} \label{aa+}
    \langle a_{\omega} \rangle_{T{\xi}} &=0\,,&\langle a_{\omega}\,a_{\omega'}\rangle_{T\xi}&=-\mu_{\omega\omega'}\nu_{\omega\omega'} (1+n_\omega+n_{\omega'}) \,,&\langle a_{\omega}^{\dagger}a_{\omega'}\rangle_{T\xi} &=n_{\omega} +\eta_{\omega \omega'}^2 (1+ 2 n_\omega^2)
 \delta(\omega-\omega')\,, \end{align} where
$\mu_{\omega\omega'}=\cosh r_{\omega\omega'}$,
$\nu_{\omega\omega'}=e^{i\theta_{\omega\omega'}}\,\sinh
r_{\omega\omega'}$ and
$\eta_{\omega\omega'}=|\nu_{\omega\omega'}|$. Notice that the
retarded Green's function~(\ref{G_HR}), given by the expectation
value of the commutator of the field $F$, remains the same in the
two-mode squeezed thermal state because the involved Bogoliubov
transformations are the canonical transformations that preserve
the commutation relations between the creation and annihilation
operators. Moreover the position correlator $\langle X(t) \, X(t')
\rangle$ in the squeezed vacuum and thermal states can be
calculated straightforwardly.

Now we are in the stage to find the corresponding Green's function
of the environment field in squeezed thermal state, which can be
obtained by the associated Langevin equation. The Langevin
equation of the mirror with the effects from the environment can
be straightforwardly derived from the influence
functional~\cite{Yeh_16_1} as
  \be
  \label{langevin_eq} \int dt' \, G_R (t, t') \, X(t')=\eta (t) \, .
  \ee
The noise force correlation function is given by
\begin{equation}
\langle \eta (t) \, \eta (t') \rangle =  \, G_H (t, t') \, . \label{langevin1}
\end{equation}
Then, according to the Langevin equation in its Fourier
transformed form, the fluctuations on the position can be
related to the retarded and Hadamard functions with respect to the
normal vacuum and thermal states as follows:
 \be \label{XX_omega}
  \langle X(\omega) X(-\omega)
\rangle= \frac{G_H  (\omega)}{G_R  (\omega) G^{ *}_R (\omega)} \,
. \ee {At finite temperature, the
Langevin equation gives the relation}
 \be \label{G_U_T} G_H^{T}
(\omega) =\pi \bigg( \frac{e^{\frac{\omega}{T}}
+1}{e^{\frac{\omega}{T}} -1} \bigg)  \, G_R^T  (\omega) G^{T *}_R
(\omega)\,  U^{T \, 2}_{\omega} (r_b)\, .
 \ee
We can check that  our approximate results for Green's functions
in~(\ref{GRT}),(\ref{modeT}), (\ref{GTH}) and (\ref{UT}), in the
low frequency limit,  satisfy this relation. A similar relation at
zero-$T$ can be found by taking the $T \rightarrow 0$ limit. Using
the Langevin equation in~(\ref{langevin_eq}) and
(\ref{langevin1}), we  find the corresponding  Hadamard function
of the boundary fields in the squeezed thermal states  is
\begin{align}\label{G_H_sT}
   { G_{H}^{(T \xi)}(t,t') }
   & = \int_{0}^{\infty}\! {d\omega}\int_{0}^{\infty}\!{d\omega'}\; {W}(\omega) {W}(\omega') \, U^T_{\omega} \, U^T_{\omega'} \, G_R^{T} (\omega) \, G^{T *}_R (\omega') \notag \\
   &  \biggl[- \mu_{\omega\omega'} \nu_{\omega\omega'}  (1+n_\omega+n_{\omega'}) \,{\frac{G_R^{T} (\omega')}{G_R^{T *} (\omega')}} \, e^{-i\omega t-i\omega' t'}\biggr.\notag\\
   &\quad\quad +\biggl. \delta(\omega-\omega')  (\eta_{\omega\omega'}^2 +\frac{1}{2} )
\, (1+2 n_\omega)  \, e^{-i\omega t+i\omega'
    t'} \biggr]+\mbox{c.c.}\, ,
\end{align}
where $G_R^{T} (\omega)$ is the Fourier transform of the retarded
Green's function  in the normal thermal state in~(\ref{GTR}). In the above
expression, we have introduced the simplest window function
${W}(\omega)$  given by the unit-step function
\begin{align}\label{w}
    {W}(\omega)&=1\,,&&\text{if $\omega_0-\Delta \leq\omega\leq\omega_0+\Delta$}\,.
\end{align}
Thus only modes within the frequency band $\omega_0 -\Delta
\leq\omega\leq\omega_0+\Delta$ are excited to the squeezed thermal
states. The other modes remain in normal thermal states. This
result can be compared to the perturbed holographic influence
functional derived in~(\ref{S_phi}). From (\ref{S_phi}) with the expression~(\ref{Q_pm}) and the influence functional~(\ref{influencefun2}), the
corrections to the Hadamard function of boundary fields in thermal
states, denoted by $G_{H}^{(\phi)}(t,t')$, can be obtained as
\be\label{G_H_phi}
    G_{H}^{(\phi)}(t,t')= \int_{0}^{\infty}\!\frac{d\omega}{2\pi}\int_{0}^{\infty} \frac{d \omega'}{2\pi}\; \, G_H^T (\omega)\,\left\{ \phi(\omega+\omega',r_b) \, e^{-i\omega t-i\omega'
    t'}+\phi(\omega-\omega',r_b) \, e^{-i\omega t+i\omega'
    t'}+\mbox{c.c.}\right\} \, .
  \ee
In the limits of small squeeze parameters and the narrow bandwidth
( $\Delta/\omega_0 <1$ in (\ref{w}) ), we can approximate $\omega
\approx \omega'$ since $\omega$ and $ \omega'$ lie within the
frequency band. Compared~(\ref{G_H_phi}) with~(\ref{G_H_sT}) and
using~(\ref{G_U_T}), the field $\varphi(2 \omega)$ obtained from
$\phi(\omega, r_b)$ in~(\ref{varphi}) can be related to the
squeeze parameters up to a constant phase by
  \be \label{dual}
  r_b^{-2-z}\varphi {(\omega+\omega' \approx 2
\omega)}= - 4 \pi \, {r_{\omega\omega}}  \, \Gamma
(\frac{3}{2}+\frac{1}{z} )\, \left(\frac{\omega}{2 z}
\right)^{-\frac{2+z}{2z}}\, .
 \ee
The large $r_b$ limit is taken to obtain (\ref{dual}). Notice that
this identification is held for any temperature, and thus, as
expected, the same identification is found at zero temperature
in~\cite{Yeh_16_1}. Also, the squeeze parameters are expected to
be small, since the holographic dual of squeezing vacuum (thermal)
states is considered in gravitational wave perturbed background.
Later, we will express some of the results in terms of general
squeezing parameters, however it should keep in mind that small
squeeze parameters are considered.

\section{Uncertainties on the position and momentum of the mirror} \label{sec3}
The presence of the environment will give additional uncertainties
to the observables associated with the mirror. The effects from
the environment on the uncertainties of the position and momentum
of the mirror are contained in the two-point functions of the
environment field. In the following we consider the environment
field to be strongly coupled and use the holographic influence
functional discussed in the previous section to study its effect
on the mirror's uncertainties in the position and momentum. For finding correlators of mirror's position, we have proven
the equivalence between influence functional method and the method of the  mode
expansion via the identification (\ref{dual}). It is then quite
straightforward to compute two-point correlation of $ X(t)$ from
the mode expansion in~(\ref{modeexpand}) with the squeezed thermal
state. We will mainly study the late time behavior of the
uncertainties, resulting from the small frequency limit of the
mode function.

\subsection{The uncertainties in squeezed vacuum states}
Let us now discuss the squeezed vacuum state by taking zero
temperature limit. Here we choose the squeeze parameter as
$\xi_{\omega \omega'}=\xi_\omega \delta(\omega-\omega')$. We also
consider that the interaction between the mirror and fields is
turned on at $t_i=-\infty$. Then the difference of the mirror's
position uncertainty at time $t$ and the result at time $t=0$ can be
expressed as
 \bea \label{deltaXuExp} \langle (X(t)-X(0)))^2
\rangle_{\xi} &&=\int_{0}^{\infty}\!{d\omega} \, {W}(\omega) \,
U_\omega^{(0) \, 2} \,
 \biggl[- \mu_{\omega}
\nu_{\omega}\,(e^{-i\omega t}-1)^2 \biggr. \nonumber
\\
&& \quad \quad\biggl. - \mu_{\omega} \nu^{*}_{\omega} \,(e^{+i\omega
t}-1)^2\biggr.+\biggl. ( 2 \, \eta_{\omega}^2+1)\, (e^{-i\omega t}-1)
(e^{+i\omega t}-1)\biggr]\,\nonumber\\
&&= \int_{\omega_0-\Delta}^{\omega_0+\Delta}\!{d\omega}\;
8 \,  U_\omega^{(0) \, 2} \, \bigg[
 \mu \, \eta \,\cos[\,\omega t\,-\theta]+(\eta^2+\frac{1}{2} )\bigg] \,
\sin^2\frac{\omega t}{2} \, ,
 \eea
where the mode function $U_\omega^{(0)}$ of the fields at vacuum
is given in~(\ref{U0}). The window function $ {W}(\omega) $ with a
finite bandwidth is also included for squeezed modes, in which the
squeeze parameters are assumed to be independent of frequency
within the frequency band. The saturated value of the uncertainty
can be found from the late-time behavior of (\ref{deltaXuExp}) in
the limit $ (\omega_0 \pm \Delta) t \gg 1$. In this limit, the
main contribution to the integration comes from the regions of
small $\omega$.  The small $\omega$ expansion of $U_\omega^{(0) \,
2}$ takes different forms for $1<z<n+2$ and $ z>n+2$, and they are
respectively given by
\begin{align}
 \label{smallw}
 U_\omega^{(0) \, 2}  & \simeq\begin{cases}
                                    \displaystyle \,\frac{\mathcal{N}_{1<z<n+2}}{T_{n+2} S_n }\,(r_b^z)^{2-{(4+2n)}/{z}}\,
                                    \omega^{-3+(n+2)/z}\,; \quad\quad \mathcal{N}_{1<z<n+2}=\frac{(2z)^{2-{(n+2)}/{z}}}{ \Gamma^2(\frac{n+2}{2 z}-\frac{1}{2})}\,, \quad 1<z<n+2\, ;& \vspace{9pt}\\
                                    \displaystyle  \,\frac{\mathcal{N}_{z>n+2}}{T_{n+2} S_n }\, \omega^{-1-(n+2)/z}\,; \quad\quad \quad\quad \mathcal{N}_{z>n+2}=\frac{(2z)^{{(n+2)}/{z}}}{ \Gamma^2(\frac{1}{2}-\frac{n+2}{2 z})}\,, \quad  z>n+2\,,&
                                \end{cases}
\end{align}
Notice that different $\omega$-dependence in these two regions of
$z$ is mainly attributed to the fact that the low frequency
behavior of the retarded Green's function in~(\ref{G0R}) is
dominated respectively by the mass term when $1 < z < n+2$ and by
the $\gamma$ term when $z > n+ 2$.

Using the small $\omega$ expansion of the mode functions in
(\ref{smallw}), the time dependence of the momentum and position
uncertainties is explored in the following. We first study the
momentum uncertainty, which is obtained directly
from~(\ref{deltaXuExp})
 {by the relation,} $P=m d X/dt$,  and the result is:
\begin{align} \label{deltaPu}
& (\Delta P (t) )^2_\xi =\langle (P(t)-P(0)))^2 \rangle_{\xi}   \\
&\simeq\begin{cases}
                                    \displaystyle \,\frac{ \mathcal{N}_{1<z<n+2}}{ T_{n+1} S_n  } \,\frac{g_{-} (r,\theta)}{4 (n+2)/z} (r_b^z)^{2-(4+2n)/z} \, m_n^2\, \bigg[(\omega_0+\Delta)^{(n+2)/z}-(\omega_0-\Delta)^{(n+2)/z}\bigg]\\
                                    \quad\quad\quad \quad\quad\quad  \quad\quad\quad \quad  \quad\quad\quad \quad\quad\quad  \quad\quad\quad \quad\quad\quad \quad\quad+\mathcal{O}(1/t)\,,
                                    \quad  \quad 1<z<n+2\, ;& \vspace{9pt}\\
                                    \displaystyle \,\frac{\, \mathcal{N}_{z>n+2}}{T_{n+1} S_n} \, \frac{g_{-} (r,\theta)}{4 (2-(n+2)/z)} \,m_n^2\, \bigg[  (\omega_0+\Delta)^{2-(n+2)/z}  -(\omega_0-\Delta)^{2-(n+2)/z} \bigg]+\mathcal{O}(1/t
                                    )\,,\\
                                    \quad\quad\quad \quad\quad\quad  \quad\quad\quad \quad\quad\quad  \quad\quad\quad \quad\quad\quad  \quad\quad\quad \quad\quad\quad \quad\quad\quad \quad\quad\quad z>n+2\,,&
                                \end{cases} \\
 &\simeq\begin{cases}
                                    \displaystyle \, {\, \mathcal{N}_{1<z<n+2}\, (T_{n+1} S_n L^2)}\,\frac{g_{-} (r,\theta)}{2} \bigg(\frac{L}{\lambda_0}\bigg)^{(n+2)/z} \, \bigg(\frac{\Delta}{\omega_0}\bigg)\,  \frac{1}{L^2}+\mathcal{O}(1/t)\,,
                                    \quad\quad 1<z<n+2\, ;& \vspace{9pt}\\
                                    \displaystyle \,{ \, \mathcal{N}_{z>n+2}\, (T_{n+1} S_n L^2)} \,\frac{g_{-} (r,\theta)}{2} \, \bigg( \frac{L}{\lambda_0} \bigg)^{2-(n+2)/z} \bigg(\frac{\Delta}{\omega_0} \bigg) \, \frac{1}{L^2}+\mathcal{O}(1/t
                                    )\,, \quad z>n+2\,.&
                                \end{cases}
\end{align}
where the function $g_{\pm}$'s of squeeze parameters are defined
as
 \be {g_{\pm}(\eta,\theta)=(2\eta^2+1) {\pm} \eta \mu
\cos(\theta)} \, .\label{gpm}
 \ee
Thus, the momentum uncertainty due to the squeezed environment
fields reach a saturated value at late times, following a
power-law saturation rate of $ t^{-1}$. The last expression is
obtained by taking the narrow bandwidth approximation ($\Delta \ll
\omega_0$) and setting $r_b^z =1/L$, which is the length scale
characterizing the breakdown of Lorentz invariance in quantum
critical theory, introduced by~\cite{Visser_09}. We also
parameterize the mass in (\ref{m_mu_n}) as $m_n= T_{n+1} S_n L$
where $1/L$ is the largest energy scale in this system. The
typical wavelength $\lambda_0\equiv 1/\omega_0$ of the squeezed
modes is in general greater than $L$. So the maximum momentum
uncertainty can be achieved when $z=n+2$. The similar results on
the velocity dispersion were also found in our earlier
work~\cite{Yeh_16_1}. It is worth noticing that the momentum
uncertainty of the mirror is proportional to the brane tension
$T_{n+1}$, which is related to the 't Hooft coupling of the
boundary field by $T_{n+1}\propto\lambda^{n/4+1/2}$. The
enhancement in the momentum uncertainty from the environment
agrees with the result in the field theory
calculations~\cite{Unruh_89}, where they also considered the
linear coupling of the environment field to particle's position
although the environment field under study is a free field. Here,
we also find that the momentum uncertainty is proportional to some
negative power of the wavelength $\lambda_0$ of the squeezed
modes. So the small value of the ratio $L/\lambda_0$ and the
narrow bandwidth approximation ($\Delta/\omega_0 \ll 1)$ can
reduce the momentum uncertainty.

Now we compute the position uncertainty. From the equation
(\ref{deltaXuExp}), the position uncertainty is given by
\begin{align} \label{deltaXu}
& (\Delta X (t) )^2_\xi=\langle (X(t)-X(0)))^2 \rangle_{\xi}   \\
&\simeq\begin{cases}
                                    \displaystyle \,\frac{\, \mathcal{N}_{1<z<n+2}}{ T_{n+1} S_n  } \,\frac{g_{+} (r,\theta)}{4 (2- (n+2)/z)} (r_b^z)^{2-(4+2n)/z} \,[(\omega_0-\Delta)^{-2+(n+2)/z}-(\omega_0+\Delta)^{-2+(n+2)/z}]\\
                                    \quad\quad\quad \quad\quad\quad  \quad\quad\quad \quad\quad\quad  \quad\quad\quad \quad\quad\quad  \quad\quad\quad \quad\quad\quad \quad\quad\quad  +\mathcal{O}(1/t)\,,  \quad 1<z<n+2\, ;& \vspace{9pt}\\
                                    \displaystyle \,\frac{ \, \mathcal{N}_{z>n+2}}{T_{n+1} S_n} \,\frac{g_{+} (r,\theta)}{4 (n+2)/z} \,\bigg[  (\omega_0-\Delta)^{-(n+2)/z}  -(\omega_0+\Delta)^{-(n+2)/z} \bigg]+\mathcal{O}(1/t
                                    )\,,\quad z>n+2\,,&
                                \end{cases} \\
 &\simeq\begin{cases}
                                    \displaystyle \,\frac{\, \mathcal{N}_{1<z<n+2}}{ T_{n+1} S_n L^2 } \,\frac{g_{+} (r,\theta)}{2} \bigg(\frac{L}{\lambda_0}\bigg)^{-2+(2+n)/z} \, \bigg(\frac{\Delta}{\omega_0}\bigg)\, L^2 +\mathcal{O}(1/t)\,,
                                    \quad\quad 1<z<n+2\, ;& \vspace{9pt}\\
                                    \displaystyle \,\frac{ \, \mathcal{N}_{z>n+2}}{T_{n+1} S_n L^2} \,\frac{g_{+} (r,\theta)}{2} \, \bigg( \frac{L}{\lambda_0} \bigg)^{-(n+2)/z} \bigg(\frac{\Delta}{\omega_0} \bigg)\, L^2+\mathcal{O}(1/t
                                    )\,, \quad z>n+2\,,&
                                \end{cases}
\end{align}
Similar to the momentum uncertainty, the saturation of the
position uncertainty is reached following a power-law behavior
like $t^{-1}$. Again, the last expression is obtained by taking
the narrow bandwidth approximation and also setting $r_b^z=1/L $.
Also, the maximum position uncertainty occurs at $z=n+2$. On
contrary to the momentum uncertainty, the position uncertainty is
inversely proportional to the brane tension $T_{n+1}$, and is
suppressed by the large coupling constant. Thus, the environment
effect reduces the position uncertainty on the one hand, and
enhances the momentum uncertainty on the other hand. Similar
environment effects from free fields on the particle are also seen
in~\cite{Unruh_89} where the interaction to the environment is via
the position of the mirror. Based on the relation $P=m dX/dt$ and
the momentum uncertainty~(\ref{deltaPu}), the position uncertainty
is proportional to some positive power of $\lambda_0$ instead.
Although the narrow bandwidth approximation reduces the position
uncertainty, the large value of the ratio $\lambda_0/L$ will lead
to some enhancement.

As also discussed in our work~\cite{Yeh_16_1}, the saturated value
of the position and momentum uncertainties depend on the functions
$g_{\pm} ( r,\theta) $ of squeeze parameters in~(\ref{gpm}). Since
 \be
 \label{g_function}\eta^2 - \frac{1}{2} \eta \mu \ge
-\frac{2-\sqrt{3}}{4} > -\frac{1}{2} \, ,
  \ee
the functions $\eta^2-\frac{1}{2} \eta \mu$ can be negative for
small squeezing parameter $r$, leading to the so-called subvacuum
phenomenon. It means that the position or momentum uncertainty,
arising from the squeezed vacuum of the environment, can be
smaller than the value solely due to the normal vacuum
fluctuations. However, the sum of the uncertainties given by the
normal vacuum and the shifted value due to squeezing vacuum must
be positive.

To fully understand environmental effects on the mirror, we study
the cross correlation between the position and momentum
uncertainties, which can be obtained straightforwardly as
\begin{align} \label{deltaXPu}
& \frac{1}{2} \big\{ \langle (X(t)-X(0))_{\xi} (P(t)-P(0))\rangle_{\xi}+ \langle (P(t)-P(0))(X(t)-X(0))_{\xi} \rangle_{\xi} \big\}  \\
&\simeq\begin{cases}
                                    \displaystyle \,- \frac{\, \mathcal{N}_{1<z<n+2}}{ T_{n+1} S_n  } \,\frac{\eta \mu \sin\theta}{4 ((n+2)/z-1)} (r_b^z)^{2-(4+2n)/z} \, m_n \, \bigg[(\omega_0+\Delta)^{(n+2)/z-1}-(\omega_0-\Delta)^{(n+2)/z-1}\bigg]\\
                                     \quad\quad\quad \quad\quad\quad  \quad\quad\quad \quad\quad\quad  \quad\quad\quad \quad\quad\quad  \quad\quad\quad \quad\quad\quad \quad\quad\quad+\mathcal{O}(1/t)\,,   \quad 1<z<n+2\, ;& \vspace{9pt}\\
                                    \displaystyle \,-\frac{ \, \mathcal{N}_{z>n+2}}{T_{n+1} S_n} \, \frac{\eta \mu \sin \theta}{4 (1-(n+2)/z)} \,m_n \, \bigg[  (\omega_0+\Delta)^{1-(n+2)/z}  -(\omega_0-\Delta)^{1-(n+2)/z} \bigg]+\mathcal{O}(1/t
                                    )\,,\\
                                    \quad\quad\quad \quad\quad\quad  \quad\quad\quad \quad\quad\quad  \quad\quad\quad \quad\quad\quad  \quad\quad\quad \quad\quad\quad \quad\quad\quad \quad\quad\quad z>n+2\,,&
                                \end{cases} \\
 &\simeq\begin{cases}
                                    \displaystyle \, {-\, \mathcal{N}_{1<z<n+2}}\,\frac{\eta \mu \sin \theta}{2} \bigg(\frac{L}{\lambda_0}\bigg)^{(n+2)/z-1} \, \bigg(\frac{\Delta}{\omega_0}\bigg) +\mathcal{O}(1/t)\,,
                                    \quad\quad 1<z<n+2\, ;& \vspace{9pt}\\
                                    \displaystyle \,{- \, \mathcal{N}_{z>n+2}} \,\frac{\eta \mu \sin \theta}{2} \, \bigg( \frac{L}{\lambda_0} \bigg)^{1-(n+2)/z} \bigg(\frac{\Delta}{\omega_0} \bigg)+\mathcal{O}(1/t
                                    )\,, \quad z>n+2\,.&
                                \end{cases}
\end{align}
The above cross correlations are found to have no dependence on
$T_{n+1}$. Since, in the holographic approach, the quadratic DBI
action in (\ref{S_DBI}) is proportional to $T_{n+1}$, the proper
rescaling of $X$ by absorbing $T_{n+1}$
 gives that $\Delta X \propto 1/\sqrt{T_{n+1}}$. Moreover,
since the mass of the mirror is proportional to the energy cost to
create the brane,  $m_n \propto T_{n+1}$~\cite{Tong_12}. As a
result, $\Delta P=m_n \Delta (d X/dt) \propto \sqrt{T_{n+1}}$.
Therefore, the product of the position and momentum uncertainties,
and so the cross correlations have no dependence on the coupling
constant of strongly coupled fields. This is probably a general
consequence from the holographic approach.

In particular, with the position and momentum uncertainties and
their correlations in ~(\ref{deltaXu}),(\ref{deltaPu}),and
(\ref{deltaXPu}) respectively, {we find that} when
$t \rightarrow \infty$,
 { \be
{(\Delta P (\infty) )^2_\xi \, (\Delta X (\infty) )^2_\xi \propto
\mathcal{N}^2 \bigg[ (\eta^2+\frac{1}{2})^2-\frac{\eta^2 \mu^2}{4}
\bigg] \bigg(\frac{L}{\lambda_0}\bigg)^{2 \vert 1-(n+2)/z \vert}
\bigg(\frac{\Delta}{\omega_0}\bigg)^2 \,,} \label{w_xi} \ee} where
$\mathcal{N}$ can be $\mathcal{N}_{1<z< n+2}$ or $\mathcal{N}_{z>
n+2}$, depending on the value of $z$. The $z$-dependence
of~(\ref{w_xi}) shows that, when $z=n+2$, $(\Delta P (t) )^2_\xi
(\Delta X (t) )^2_\xi$ is largest for $L <\lambda_0$ as compared
with other $z$. Thus, at $z=n+2$, the environment effects on the
mirror is maximal.

Accordingly, for
considering the same frequencies of the squeezed modes and
squeezing parameters, the quantum critical theories with the
dynamical exponent $z=n+2$ gives maximum uncertainty effects on
mirror's position and momentum. Also, the position uncertainty can
be reduced by the large coupling constant of the strongly coupled
fields, whereas the momentum uncertainty is enhanced  by the
coupling constant. It deserves a further study on finding the
holographic dual of the system-environment model, where the
interaction between them is via system's momentum, to explore the
dependence of the position and momentum uncertainties on the
coupling constant of the environment fields.

\subsection{The uncertainties in squeezed thermal states}

As for the environment in squeezed thermal state, the mirror can
receive the significant finite-$T$ effects. This can be seen from
the retarded Green's function in the small frequency
limit~(\ref{GTR}), which gives finite-$T$ modification to the mass
 and dissipation coefficient $\gamma_T$.
On top of that, thermal fluctuations of the environment,
summarized in the Hadamard function~(\ref{GTH}), renders the
mirror undergoing stochastic motion.  So the uncertainties of the
position and momentum of the mirror are modified when the
environment is heated up.

Using the mode functions at finite-$T$ and the expectation values
of creation and annihilation operators for squeezed thermal
states~(\ref{aa+}),  we find the position uncertainty as,
  \bea
\label{deltaXuTExp} && \langle (X(t)-X(0)))^2 \rangle_{T\xi}
=\int_{0}^{\infty}\!{d\omega}\, {W}(\omega) \,  U_\omega^{T \, 2}
\,
 \biggl[{-} \mu_{\omega}
\nu_{\omega}\, (1+ 2 n_{\omega}) \,(e^{-i\omega t}-1)^2 \biggr. \nonumber
\\
&&  \quad\quad\quad \quad\quad\quad\biggl. {-} \mu_{\omega} \nu^{*}_{\omega} \,(1+2 n_{\omega}) \, (e^{+i\omega
t}-1)^2\biggr.+\biggl. ( 2 \, \eta_{\omega}^2+1)\, (1+2 n_{\omega}) \, (e^{-i\omega t}-1)
(e^{+i\omega t}-1)\biggr]\,\nonumber\\
&&\quad\quad\quad= \int_{\omega_0-\Delta}^{\omega_0+\Delta}\!{d\omega}\;\,
8 \,  U_\omega^{T \, 2} \, \bigg[
 \mu \, \eta \,(1+ 2 n_{\omega}) \, \cos[\,\omega t\,-\theta]+(\eta^2+\frac{1}{2})(1+2 n_{\omega}) \bigg] \,
\sin^2\frac{\omega t}{2} \, ,\nonumber \\
 \eea
where $U_{\omega}^{T 2}$ is the square of mode functions
in~(\ref{UT}). The squeeze parameters are assumed to be
 $\xi_{\omega \omega'}=\xi_\omega \delta (\omega-\omega')$ and frequency-independent in the frequency
band specified by the window function $W$. However the mode
functions in~(\ref{UT}) can be found only in the low-frequency
limit (or equivalently high temperature limit), so we will
restrict our study in this limit. Thus, as long as the squeezed
modes under consideration have frequency $\omega \ll T$, the
number density $n_\omega$ can be approximated by $n_\omega \approx
T/\omega$.

As a result, the momentum uncertainty given by the squeezed
thermal environment at high-$T$ limit ($T >> \omega_0$) becomes
 \bea \label{deltaPuT}
&& (\Delta P (t) )^2_{T\xi}=\langle (P(t)-P(0)))^2 \rangle_{T\xi} \nonumber\\
&&\simeq \frac{8\, }{\pi T_{n+1} S_n} m^2_{n T} \, \omega_0^2 \, r_h^{-n-2} \,\bigg( \frac{T}{\omega_0}\bigg)\, g_{+} (r,\theta) \, \bigg(\frac{\Delta}{\omega_0}\bigg)+\mathcal{O}(1/t) \nonumber\\
&&\propto {\,  (T_{n+1} S_n L^2)} \,
\bigg(\frac{L}{\lambda_0}\bigg)^2 \,
\frac{T^{-(n+2)/z}}{L^{(n+2)/z}} \,\bigg( \frac{T}{\omega_0}
\bigg)\, g_{+} (r,\theta) \, \bigg(\frac{\Delta}{\omega_0}\bigg)
\frac{1}{L^2}+\mathcal{O}(1/t)  \, .
 \eea
In particular, the last expression is obtained by assuming that
$T$ is larger than the frequency $\omega_0$ of the squeezed modes,
but is still smaller than $1/L$, the largest energy scale in this
system. If so, the mass $m_{n T}$ in~(\ref{m_gamma_T}) can be
approximated by $m_{nT} \simeq T_{n+1} S_n/L$ with no temperature
dependence. Similar to the results in the zero temperature case,
the momentum uncertainty reaches its saturated value following a
power-law $1/t$, and is enhanced by the factor of brane's tension
$T_{n+1}$.


It may be quite instructive if the above dependence of the
momentum uncertainty on temperature can be reconstructed by
dimensional analysis using the energy scales $T$ and
$\omega_0$~\cite{Tong_12}. Here we take a Brownian particle as an
example. The same arguments will also be applied to a
n-dimensional mirror by shifting the value of $z$ from $z=2$ to
$z=n+2$. The mean free path of a particle can be argued to be
$\ell_{mfp} \propto 1/T^{1/z} $ due to the scaling symmetry of
quantum critical theories in~(\ref{E:ernfd}). Moreover, the relaxation
time is inversely proportional to $T$ as $\tau \propto 1/T$. For
the typical measuring time $t=1/\omega_0$, the relevant time scale
given by the squeezed modes, the number of collisions is
approximated by $N_{c}=t/\tau \propto T/\omega_0$. Therefore,
$\Delta X^2 =N_c \ell_{mfp}^2 \propto T^{1-2/z}/\omega_0$, as will
also be seen later by direct calculations. The corresponding
momentum uncertainty, from the relation $P=m dX/dt$ and taking the
relevant time scale of the mirror $t=1/\omega_0$, becomes $\Delta
P^2 \propto T^{1-2/z} \omega_0$, which for a $n$-dimensional
mirror is modified to $\Delta P^2 \propto T^{1-(n+2)/z} \omega_0$
as above. However it is peculiar
that for $1<z< n+2$, the momentum uncertainty is inversely
proportional to $T$. This unanticipated result can be tested
experimentally in the future.~\cite{Holographic QBM,
Tong_12,Hartnoll_10}.

Now we turn to the corresponding position uncertainty, which is
obtained as
 \bea \label{deltaXuT}
&&(\Delta X (t) )^2_{T\xi}=\langle (X(t)-X(0)))^2 \rangle_{T\xi}\nonumber\\
 &&\simeq \frac{4\, }{\pi T_{n+1} S_n}  \, r_h^{-n-2} \,\omega_0\, \bigg( \frac{T}{\omega_0}\bigg)\, g_{-} (r,\theta) \, [ (\omega_0-\Delta)^{-1}-(\omega_0+\Delta)^{-1}] +\mathcal{O}(1/t) \nonumber\\
&&\propto \frac{1}{T_{n+1} S_n L^2} \,
\frac{T^{-(n+2)/z}}{L^{(n+2)/z}} \,\bigg( \frac{T}{\omega_0}
\bigg)\, g_{-} (r,\theta) \, \bigg(\frac{\Delta}{\omega_0}\bigg)
\, L^2  +\mathcal{O}(1/t)\, .
 \eea
As anticipated, the position uncertainty is inversely proportional
to the brane's tension $T_{n+1}$ and its temperature dependence
shares the same behavior as in the momentum uncertainty. Finally,
the cross correlation between the momentum and position
uncertainties is
 \bea \label{deltaPXuT}
 &&\frac{1}{2} \big\{ \langle (X(t)-X(0)) (P(t)-P(0))\rangle_{T\xi}+ \langle (P(t)-P(0))(X(t)-X(0)) \rangle_{T\xi} \big\}  \nonumber \\
 && \quad\quad\simeq -\frac{4\, }{\pi T_{n+1} S_n} m_{nT} \, \omega_0 \, r_h^{-n-2} \,\bigg( \frac{T}{\omega_0}\bigg)\, \mu\eta \sin\theta \, \ln \bigg[ \frac{\omega_0+\Delta}{\omega_0-\Delta} \bigg] +\mathcal{O}(1/t)\nonumber\\
&& \quad\quad \propto  - \,
\bigg(\frac{L}{\lambda_0}\bigg)\,\frac{T^{-(n+2)/z}}{L^{(n+2)/z}}
\,\bigg( \frac{T}{\omega_0} \bigg)\, \mu \eta \sin\theta \,
\bigg(\frac{\Delta}{\omega_0}\bigg)+\mathcal{O}(1/t) \, .
 \eea
As can be seen, the cross correlation has the same temperature
dependence as in the position and momentum uncertainties, and also
it has no $T_{n+1}$ dependence. To sum up,
 the corresponding product of the position and momentum
uncertainties when $t\rightarrow \infty$ is
 \be \label{w_Txi}
{(\Delta P (\infty) )^2_{T\xi} \, (\Delta X (\infty) )^2_{T\xi}
\propto \frac{256}{\pi^2} \, \bigg[
(\eta^2+\frac{1}{2})^2-\frac{\eta^2 \mu^2}{4}   \bigg]\, \big( T
/L^{-1} \big)^{2 (1-(n+2)/z)} \,
\bigg(\frac{\Delta}{\omega_0}\bigg)^2 }\, ,
 \ee
in the high-$T$ ($T >> \omega_0$) and narrow bandwidth limits.
Thus, for $1<z<n+2$, the contribution from the fluctuation of the
squeezed thermal states to the product of the position and
momentum uncertainties decreases as $T$ increases whereas for
$z>n+2$, the contribution increases instead as $T$ increases.
The peculiar temperature dependence seems to be the
consequence from  the scaling symmetry of quantum critical
theories under study, and deserve an experimental check.

\subsection{Comparison with the case in the environment of the relativistic free-field}
The effects on the system from the strongly coupled environment
field we obtain can be compared with the results from free field
theories. We concentrate on the cases with the dynamical exponent
$z=1$, and consider the system consisting of a particle, thus
$n=0$, since these cases are what have been studied in field
theories. In~\cite{FB}, they consider a bilinear coupling between
the particle position and the relativistic free field with a
coupling constant $\epsilon$. The uncertainties of particle's
momentum and position, which arise from quantum and thermal
fluctuations of the free relativistic fields {with all frequency
modes}, are calculated. In particular, the saturated value of the
momentum uncertainty for a free particle affected by the thermal
bath in high-temperature $T$ limit is found to be $\langle \Delta
P (t\rightarrow\infty) \rangle^2_T \propto m \, \epsilon^2 T $,
where $m$ here is the mass of a particle. Thus, the increase in
$\epsilon$ gives rise to larger momentum uncertainty~\cite{FB}. On
the contrary, the position uncertainty in the late time limit does
not saturate and they found, $\langle \Delta X (t) \rangle^2_T
\propto (\epsilon^2 T /m \gamma) t $, where $\gamma$ is the
damping constant. So the position uncertainty increases with the
square root of time just as in the case of classical Brownian
motion. However, the damping constant $\gamma$ is obtained from
the retarded Green's function of the field, which is proportional
to the coupling constant $\epsilon^2$, and is independent of
temperature $T$~\cite{FB}. As a result, the position uncertainty
is independent of the coupling constant $\epsilon$, and both
position and momentum uncertainties increases linearly in
temperature $T$ of the environment. As for the system interacting
with the strong coupled field under consideration, there is an
additional large coupling constant of the quantum critical fields,
relative to the interaction strength between the system and the
environment, the effects from the environment to the system are
mainly determined that large coupling constant of the fields
instead. New features we find here is that the large coupling
constant of the field reduces the position uncertainty of the
particle, but enhances the momentum uncertainty. Moreover, in the
case $z=1$, both position and momentum uncertainties of the
particle decrease as the temperature of the heat bath increase.
This is in a sharp contrast to the behavior in the free field heat
bath. We also find that the coupling of the system to the squeezed
state of the environment leads to squeezing the quantum state of
the system itself through their bilinear coupling. Also, when
squeezed modes are restricted to some finite range of frequency,
both position and momentum uncertainties of the particle at late
times reach their respective saturated value by following the
relaxation behavior as $1/t$. Similar saturation behavior is also
found when the particle interacts with relativistic free fields in
their squeezed vacuum states with a finite frequency bandwidth
in~\cite{Lee_12}. Additionally, it has been emphasized in previous
sections that in \cite{Unruh_89}, the bilinear coupling between
particle's position and the relativistic free fields  leads to the
saturation of position and momentum uncertainties at late times
for the particle of an oscillator, and also find the reduction in
the position uncertainty in comparison with the momentum
uncertainty. With the same type of the coupling, the same
reduction behavior is found in our holographic setup to consider
the strongly coupled quantum critical fields. Thus, there exist
some dramatic differences in the effects on the system from the
environments of a free field and a strongly coupled field, and
they can be experimentally compared. These results also show that
the previous studies on open quantum systems based on perturbative
methods to deal with the weakly coupled environment fields are not
as robust as asserted even qualitatively in the case of  the
strongly coupled fields.

Brief discussions are given below for the environment consisting
of free Lifshitz field theories with the Lorentz symmetry breaking
dispersion relation in the case of a general
$z$~\cite{Visser_09,ALEX}. Although such system-environment
problems have not been studied yet, we may still expect some
differences. The same type of the bilinear coupling between the
system and free Lifshitz fields might lead to rather different
feature of the retarded Green's function and the Hadamard function
defined by~(\ref{G_HR}). In particular, the retarded Green's
function is constructed from the expectation value of the
commutator of the field variable. As for the free field, based
upon the canonical quantization, the retarded Green's function is
found to be independent of the state of the field. For example,
for the thermal state the retarded Green's function has no
temperature dependence. This is in sharp contrary to the case of
the strongly coupled field with the temperature dependent retarded
Green's function~(\ref{GTR}). As a result, the Hadamard function
of the free field, as can be seen from the fluctuation-dissipation
theorem, also has different temperature dependence, as compared
with the one in the strongly coupled field~(\ref{GTH}).
Accordingly, we also anticipate the dramatically different
temperature dependence of the momentum and the position
uncertainties, influenced by the thermal state of the free
Lifshitz field, and it deserves further study.

\section{Summary and outlook}\label{sec4}
The main goal of this work is to understand the effects of the
strongly coupled quantum critical fields on the dynamics of a
$n$-dimensional mirror using the method of holography. The dual
description is a $n+1$-dimensional probe brane moving in
$d+1$-dimensional Lifshitz geometry. The dynamics of the mirror
can be realized from the motion of the brane at the boundary of
the bulk.  The correlators of strongly coupled environment fields
at squeezed vacuum and thermal states can be obtained via
holographic influence functional, constructed from the probe brane
action in the gravitational wave perturbed Lifshitz (black-hole)
geometry. The interaction between the mirror and the environment
is a bilinear coupling through the mirror's position. We find that
the position uncertainty of the mirror due to the presence of the
environment is suppressed by the large coupling constant of the
fields but the momentum uncertainty is enhanced by the coupling
constant instead. As a result, the product of the position and
momentum uncertainties is independent of the coupling constant.
This finding can be one of the general consequences in the
holographic description of Brownian motion. The amplitude of
squeeze parameter $\eta$, counting the number of the quanta in
squeezed modes, gives additional enhancement to the uncertainties
whereas its phase factor $\theta$ may reduce the uncertainties by
some proper tuning. In the squeezed vacuum state, the mirror gains
maximum effects on its position and momentum uncertainties from
the environment when the dynamic exponent $z=n+2$. For the
squeezed thermal state, the contribution to the uncertainties from
the thermal fluctuation decreases as $T$ increases for $1<z<n+2$,
whereas for $z>n+2$ the contribution increases as $T$ increases
instead.  All results deserve experimental tests in physical
systems to justify success in employing holographic ideas for the
study of environmental effects of strongly coupled fields on the
system.

The interaction between the system and environment may result in
the loss of quantum coherence of the system. A quantitative way to
characterize the decoherence is by the entanglement entropy,
defined as $S=-Tr\rho_r\ln\rho_r$, where $\rho_r$ is the reduced
density matrix of the system. A direct extension of the current
work is to calculate the time-dependent entanglement entropy of
the system via the holographic influence functional approach when
turning on the interaction between the system and environment at
some initial time, and also imposing a suitable initial density
matrix of the system. Another possible extension is to consider
two quantum systems, coupled to one strongly coupled quantum field.
In particular, we may explore the development of their quantum
entanglement through the interaction with the strongly coupled
field. To do so, one needs to extend the current holographic setup
to include two objects moving in the asymptotic Lifshitz
background. In this case, it will be interesting to compare the
time-dependent entanglement entropy of two sub-systems derived
from the holographic influence functional approach, which is valid
in the linear response region, with the entanglement entropy obtained by the
Ryu-Takayanagi Conjecture \cite{Ryu}.

\begin{acknowledgments}
 This work was supported in part by the
Ministry of Science and Technology, Taiwan.
\end{acknowledgments}

\section*{Appendix}
\appendix

 \section{Review of the method of  influence functional}
 \label{inf}
In this appendix, we give a brief review of the method of
influence functional in field theory~\cite{Leggett,SK,Fv}.  We
begin with the total density matrix $\rho(t)$ of the
system-plus-environment, which unitarily evolves according to
\begin{equation}
 \rho (t_f) = U(t_f, t_i) \, { \rho} (t_i) \, U^{-1} (t_f,
t_i )
\end{equation}
where $ U(t_f,t_i) $ is the time evolution operator.  The effects
from the environment to the system can be summarized in the
reduced density matrix $\rho_r(t)$, obtained by tracing out the
environmental degrees of freedom in $\rho(t)$. Here the initial
total density matrix at time $t_i$ is assumed to be factorized as
 \begin{equation}
 \label{initialcond}
    \rho(t_i)=\rho_{q}(t_i)\otimes\rho_{{F}}(t_i)\, ,
 \end{equation}
where $q$ and $F$ generically represent the system and the
environment variables respectively. The environment field is
assumed initially in thermal equilibrium at temperature $T=1/\beta
$, and the corresponding  density matrix $\rho_{{F}}(t_i)$ is
given by
\begin{equation}\label{initialcondphi}
    \rho_{F}(t_i)=\rho_T \equiv e^{-\beta H_{F}}/Tr_{{F}}\{e^{-\beta H_{F}}\} \,,
\end{equation}
where $H_{F}$ is the Hamiltonian for the $F$ field. The system and
environment start to couple at an initial time $t_i$. The vacuum
state of the environment field can be achieved by taking the
zero-$T$ limit.

In the spirit of linear response, the system is considered to be
linearly coupled to the environment.  Thus, the full Lagrangian
takes the form
  \be
   L(q,F)=L_q[q]+L_F[F]+qF\, .
  \ee
One can then {express} the reduced density matrix
as~\cite{Leggett,SK,GSI}
\begin{equation}
\rho_r({q}_f,\tilde{{q}}_f,t_f)=\int\!d{q}_1\,d{q}_2\;\mathcal{J}({q}_f,\tilde{{q}}_f,t_f;{q}_1,{q}_2,t_i)\,\rho_{q}({q}_1,{q}_2,t_i)\,,\label{evolveelectron}
\end{equation}
where the propagating function
$\mathcal{J}({q}_f,\tilde{{q}}_f,t_f;{q}_1,{q}_2,t_i)$  carries
the information about the effects from the environment, and can be
expressed in terms of  the influence functional
$\mathcal{F}[{q}^+,{q}^-]$ by
\begin{equation}\label{propagator}
    \mathcal{J}({q}_f,\tilde{{q}}_f,t_f;{q}_1,{q}_2,t_i)=\int^{{q}_f}_{{q}_1}\!\!\mathcal{D}{q}^+\!\!\int^{\tilde{{q}}_f}_{{q}_2}\!\!\mathcal{D}{q}^-\;\exp\left[i\int_{t_i}^{t_f}dt\left(L_{q}[{q}^+]-L_{q}[{q}^-]\right)\right]\mathcal{F}[{q}^+,{q}^-]\,.
\end{equation}
Up to the quadratic order in particle position $q$, the influence
functional in terms of real-time Green's functions for the
environment field can be written as~\cite{Fv}
\bea\label{influencefun2}
{\mathcal{F}}\left[{q}^{+},{q}^{-}\right]&& = \exp\bigg\{
-\frac{i}{2}\int_{t_i}^{t_f} dt\!\!\int_{t_i}^{t_f} \!dt' \Big[
{q}^+(t)\,G^{++}(t,t') \,{q}^+(t')\Bigr.\bigr.
-{q}^+(t)\,G^{+-}(t,t')\,{q}^-(t') \nonumber\\
&&- {q}^-(t)\,G^{-+}(t,t')\,{q}^+(t')
\big.\big.+{q}^-(t)\,G^{--}(t,t')\,{q}^-(t')\big]\bigg\}\,.
 \eea
 The time-ordered, anti-time-ordered Green's
functions and Wightman functions are defined respectively by
    \bea
\label{correlator}
    && i\,G^{+-}(t,t')=\langle F(t')F(t)\rangle \, , \nonumber\\
    && i\,G^{-+}(t,t')=\langle F(t)F(t')\rangle \, ,\nonumber\\
    &&i\,G^{++}(t,t')=\langle F(t)F(t')\rangle\theta(t-t')+\langle
    F(t')F(t)\rangle\theta(t'-t)\, , \nonumber\\
    && i\,G^{--}(t,t')=\langle F(t')F(t)\rangle\theta(t-t')+\langle
    F(t)F(t')\rangle\theta(t'-t) \, .
    \eea
The retarded Green's function and Hadamard function, which account
for dissipative and stochastic effects on the dynamics of the
system can be constructed out of the above Green's functions by
 \bea
\label{G_HR}
    G_{R} (t-t')&\equiv & -i \theta (t-t')\langle [F(t), F(t')] \rangle = \bigg\{ G^{++} ( t,t') -G^{+-} (t,t') \bigg\}\,, \nonumber  \\
    G_{H} (t-t')&\equiv & \frac{1}{2} \langle \{ F(t), F(t') \} \rangle=\frac{i}{4} \bigg\{ G^{++} ( t,t') +G^{+-} (t,t') + G^{--} ( t,t') +G^{-+} (t,t')
    \bigg\}\,.
 \eea
When the environment respects time-translation invariance, the
Fourier transform of its Green's function can be defined as
 \begin{equation} G (\omega)=\int  d \tau \, G
(\tau) \, e^{+i \omega \tau} \, ,
 \end{equation}
and the Fourier transforms of the Green's functions, defined in
\eqref{correlator}, are given by
  \bea
  \label{SKs}
  G^{++}(\omega)&=&{{\rm Re}}G_R(\omega)+(1+2n_\omega)\,i\,{\rm Im}G_R(\omega)\, ,\nonumber\\
  G^{--}(\omega)&=&-{\rm Re}G_R(\omega)+(1+2n_\omega)\,i\,{\rm Im}G_R(\omega)\, , \nonumber\\
  G^{+-}(\omega)&=& 2n_\omega\,i\,{\rm Im}G_R(\omega)\, ,\nonumber\\
  G^{-+}(\omega)&=& 2(1+n_\omega)\,i\,{\rm Im}G_R(\omega)
  \,
  \eea
with $n_\omega=(e^{\frac{\omega}{T}}-1)^{-1}$. Notice that  the
fluctuation-dissipation relation is satisfied, and given by
 \be
 \label{FD} G_H(\omega)=- (1+2n_\omega) \, {\rm Im} G_R
(\omega) \, .
  \ee

 \section{Brief summary of the results in pure Lifshitz geometry and Lifshitz black hole}
 \label{zero order}
 Consider the Lifshitz black hole background in (\ref{lifshitz bh
 gw})without gravitational wave perturbation
\be
  \label{lifshitz bh}
  ds^2=-r^{2z}f(r)dt^2+\frac{dr^2}{f(r)r^2}+r^2 dx_i dx_i \, .
   \ee
With the same notations and assumptions as in the main text, the
DBI action for the $n+1$-dimensional probe brane in the Lifshitz
black hole for small $X^I$ is given by \be
  S^T_{DBI}
  \approx {\rm constant}- \frac{T_{n+1}}{2}\int dr \, dt \, dx_1 \, dx_2 \, ... \, dx_n \,
\bigg( r^{z+n+3} f(r) X'^{I} X'^{I}-
\frac{\dot{X}^{I}\dot{X}^{I}}{ { f(r) r^{z-n-1}}}\bigg) \, .
  \label{s_T_dbi}
  \ee
Thus the equation of motion for brane's position in the Fourier
space, $X^I_\omega (r)e^{-i\omega t}$ can be derived as follows
 \be
  \label{mirror n with T}
 \frac{\partial}{\partial r}\biggl( r^{z+n+3}f(r)\frac{\partial }{\partial
 r}  X^I_\omega (r) \biggr)+\frac{\omega^2}{r^{z-n-1}f(r)} X^I_\omega(r) =0 \,
 .
  \ee
The solution can be expressed in terms of two linearly independent
solutions with the properties $\mathcal{X}_{\omega}(r)_{\substack{
   \propto \\
   r\rightarrow r_h
  }} e^{+i\omega r_*}$ and
$\mathcal{X}^{*}_{\omega}(r)_{\substack{
   \propto \\
   r\rightarrow r_h
  }} e^{-i\omega r_*}\, ,$
where $r^*=\int drf(r)^{-1}r^{-z-1}$, and the normalization
condition $\mathcal{X}_{\omega}(r_b)=1$. Since the different
components of $X^I_\omega$ are decoupled in the linearized
equation of motion, we may just focus on one of the directions
$X^I$ and denote it by $Q(t,r)$. As described in the main text we
introduce $Q^+(t,r_1)$ and $Q^-(t,r_2)$, which correspond to the
branes living in two regions with different asymptotic boundaries
in the maximally extended Lifshitz black hole geometry.
Following~\cite{Yeh_14},
 which is  consistent with~\cite{Son_09, Son_02}, we find  $Q^{\pm} (\omega,r)$ to be
    \bea \label{Q_pm}
    &&Q^+(\omega,r_1)=\frac{1}{1-e^{-\frac{\omega}{T}}} \bigg[ (X^-(\omega)- e^{-\frac{\omega}{T}}X^+(\omega))
    \mathcal{X}_{\omega}(r_1)+ ( X^+(\omega)-X^-(\omega))
 \mathcal{X}_{\omega}^*(r_1) \bigg]\, ,\nonumber\\
    &&Q^-(\omega,r_2)= \frac{1}{1-e^{-\frac{\omega}{T}}} \bigg[ (X^-(\omega)- e^{-\frac{\omega}{T}}X^+(\omega))
    \mathcal{X}_{\omega}(r_1)+ e^{-\frac{\omega}{T}} ( X^+(\omega)-X^-(\omega))
 \mathcal{X}_{\omega}^*(r_1) \bigg] \, .  \eea
{where $X(\omega)$ is the Fourier transform of $X(t)$, the displacement of the mirror.} This
solution is then substituted into the classical on-shell
action~(\ref{S_T_DBI_onshell}). Using (\ref{influencefun2}) and
(\ref{G_HR}), the retarded Green's function at finite temperature
is obtained as
   \be
   \label{GRT}
{G_R(\omega)=T_{n+1} S_n  \, {r_b^{z+n+3}} \,
\mathcal{X}_{-\omega}(r_b)\partial_r\mathcal{X}_{\omega}(r_b)}\,.
  \ee
In general, the exact solution of $\mathcal{X}_{\omega}(r)$ at
finite temperature, denoted as $\mathcal{X}^T_{\omega}(r)$, is not
available, and depends on the details of black hole metric
(\ref{lifshitz bh}). However in the small $\omega$ limit, the
solution of $\mathcal{X}^T_{\omega}(r)$ can be derived by matching
the solution near the black hole horizon at $r=r_h$ to the
solution in the large value of $r$ with proper boundary conditions
in two regions. We then can obtain the approximate solution as
~\cite{mirror},
 \bea \label{modeT}
\mathcal{X}^T_{\omega}(r) &=& \mathcal{Y}_{\omega} (r)/ \mathcal{Y}_\omega (r_b) \, , \nonumber\\
\mathcal{Y}_{\omega} (r) &=& \frac{i}{z+n+2} \frac{\omega
  r_h^{n+2}}{
  r^{n+2+z}}\biggl[1-\frac1{\frac{n+2}{2z}+\frac32}\big(\omega/2zr^z\big)^2+\mathcal{O}(\omega^4)\biggr]
\nonumber\\
&& \quad \quad +  (1-i\,\omega
  r_h^{n+2}\kappa )
  \biggl[1+\frac1{\frac{n+2}{2z}-\frac12}\big(\omega/2zr^z\big)^2+\mathcal{O}(\omega^4)\biggr]
   \, .
 \eea
Then the retarded Green's function at finite temperature,
$G^T_R(\omega)$, in the small $\omega$ limit, can be {found as}
 \be \label{GTR}
 G^T_R (\omega)= m_{n T}(z) (i\omega)^2-\gamma_{n T}(z) (i\omega)+   \mathcal{O}(\omega^3) \,,
 \ee
where
 \be \label{m_gamma_T} m_{n
T}(z)=\frac{T_{n+1}S_n}{r_b^{z-n-2}}\biggl\{
\frac1{n+2-z}+\bigg(\frac{r_h}{r_b}\bigg)^{2n+4}\Bigl[(n+2+z)-\kappa
r_b^{z+n+2}\Bigr]\biggr\}\, , \, \gamma_{n T}(z)=T_{n+1}S_n
r_h^{n+2} \,,
 \ee
where $\kappa$ is a constant of integration. The
 mass $m_{ n T}$ and the damping
coefficient $\gamma_{nT}$ have the temperature dependence through
{their dependence on the black hole horizon radius}~(\ref{BHT}).
The peculiar dependence of $\gamma_{nT}$ on
temperature~\cite{Holographic QBM,Tong_12,Hartnoll_10},
 will play an important role in
determining the temperature effects on the position and momentum
uncertainties of the mirror to be explored later. All other
correlators can be derived from~(\ref{SKs}). In particular,
through the fluctuation-dissipation relation~(\ref{FD}), we find
the finite temperature Hadamard function
 \be
 \label{GTH}
 G^{T}_H(\omega)=\bigg( \frac{e^{\frac{\omega}{T}} +1}{e^{\frac{\omega}{T}} -1} \bigg) \, \omega \, \gamma_{n T}(z)
  \, .
 \ee

In the zero temperature limit, there is the exact expression for
$\mathcal{X}_{\omega}(r)$~\cite{mirror},
  \be
  \label{mode}
  \mathcal{X}_{\omega}(r)=\frac{r_b^{\frac{z+n+2}2}}{r^{\frac{z+n+2}2}}\frac{H^{(1)}_{\frac{n+2}{2z}+\frac{1}{2}}(\frac{\omega}{zr^z})}{H^{(1)}_{\frac{n+2}{2z}+\frac{1}{2}}(\frac{\omega}{zr_b^z})}
  \, .
  \ee
Hence the zero-temperature retarded Green's function {for
$\omega>0$} can be found
 to be,
  \be
\label{G_R} G^{(0)}_R(\omega)=- T_{n+1} S_n \,  {\omega \,
r_b^{n+2}}\frac{H^{(1)}_{\frac{n+2}{2z}-\frac12}(\frac{\omega}{zr_b^z})}{H^{(1)}_{\frac{n+2}{2z}+\frac12}(\frac{\omega}{zr_b^z})}
\,.
  \ee
Thus, in the small $\omega$ expansion,
 \be \label{G0R}
  G_R^{(0)} (\omega)={m_n (z) (i\omega)^2+\mu_n (\omega,z)} \,
  ,
  \ee
 where
  \be \label{m_mu_n}
  m_n (z)=\frac{T_{n+1} S_n}{(n+2-z)r_b^{z-n-2}},\qquad\qquad\mu_n(\omega,z)=\gamma_n(z)(-i\omega)^{1+\frac{n+2}z}+\delta_{n}(z) (-i\omega)^4+...
  \ee
with
 \be \label{gamman} \gamma_n (z)=\frac{T_{n+1}
S_n}{(2z)^{(n+2)/z}}\frac{\Gamma(\frac12-\frac{n+2}{2z})}{\Gamma(\frac12+\frac{n+2}{2z})}
\, , \qquad\quad\delta_n(\omega, z)=-\frac{T_{n+1}
S_n}{(n+2-3z)(n+2-z)^2r_b^{3z-n-2}} \, .
 \ee
The low-frequency expansion is valid as long as $\omega <\left
\vert [z-(n-2)] [z-(n+2)/3]\right\vert r_b^z$.  As mentioned
in~\cite{Tong_12}, although both $m$ and $\gamma$ change signs at
$z=n+2$, their ratio $\gamma /m$ still gives sensible results for
describing the dynamics of the mirror. The zero-temperature
Hadamard function for $\omega
>0$ is derived as
 \be \label{G_0_H}
 G^{(0)}_H(\omega)=\frac{2 z}{\pi} r_b^{n+2+z} \frac{T_{n+1} S_n} {J^2_{\frac{n+2}{2z}+\frac12}(\frac{\omega}{zr_b^z})+Y^2_{\frac{n+2}{2z}+\frac12}(\frac{\omega}{zr_b^z})}
  \, . \ee

\end{document}